\documentclass[aps, prx, reprint, superscriptaddress, 10pt]{revtex4-2}

\UseRawInputEncoding 

\usepackage{graphicx} 
\usepackage{dcolumn}  
\usepackage{bm}       
\usepackage{color}    
\usepackage{amsmath}  
\usepackage{amssymb}  
\usepackage{braket}   
\usepackage{mathrsfs} 

\usepackage{algorithm}
\usepackage{algpseudocode}

\algrenewcommand\algorithmicrequire{\textbf{Input:}}
\algrenewcommand\algorithmicensure{\textbf{Output:}}
\algnewcommand\algorithmicresources{\textbf{Resources:}}
\algnewcommand\Resources{\item[\algorithmicresources]}

\usepackage{amsthm}
\newtheoremstyle{mydefstyle}
  {\topsep} 
  {\topsep} 
  {\itshape} 
  {} 
  {} 
  {} 
  {.5em} 
  {\bfseries{\thmname{#1}\thmnumber{ #2}.}\thmnote{{ \mdseries\itshape[#3]}}} 

\theoremstyle{mydefstyle}
\newtheorem{definition}{Definition} 

\usepackage{tcolorbox} 
\usepackage{makecell}  
\usepackage{booktabs}  
\usepackage{array}     
\usepackage{multirow}  

\usepackage{lmodern}   
\usepackage{relsize}   

\usepackage{hyperref}
\hypersetup{
    colorlinks=true,
    linkcolor=blue,
    citecolor=blue,
    urlcolor=blue
}

\begin{document}

\title{Weakly-Driven Quantum Walks for Memory-Constrained Pauli Channel Learning}

\author{Yuan-Zhuo Wang}
\author{Yi-Ran Xiao}
\author{Ming-Yang Li}
\author{Shengjun Wu}\email{sjwu@nju.edu.cn}
\author{Zeng-Bing Chen}\email{zbchen@nju.edu.cn}
\affiliation{National Laboratory of Solid State Microstructures and School of Physics, Collaborative Innovation Center of Advanced Microstructures, Nanjing University, Nanjing 210093, China}

\date{\today}

\begin{abstract}
Accurate characterization of quantum noise, exemplified by the Pauli channel, is a cornerstone for building fault-tolerant quantum computers. A recent protocol (\href{https://link.aps.org/doi/10.1103/PRXQuantum.6.020323}{PRX Quantum \textbf{6}, 020323 (2025)}) combining channel concatenation and quantum memory has achieved an exponential reduction in measurement complexity for Pauli channel estimation. This efficiency, however, hinges on using logarithmic quantum memory to suppress hypothesis test errors. In this work, we introduce a mechanism termed the ``weakly-driven quantum walk'' to mitigate the demand for high-quality quantum memory. By exploiting the distinct dynamical properties of quantum walks under biased versus unbiased driving, our algorithm lowers the quantum memory overhead to a constant order while preserving the exponential advantage in measurement complexity. By analogy with weak measurement, our introduced concept of ``weak driving'' preserves pointer coherence even when driven by classical probabilistic information, a principle that may inspire new approaches to similar quantum algorithm design and quantum sensing of weak signals in resource-constrained scenarios.
\end{abstract}

\maketitle

\section{Introduction}
\label{sec:introduction}

Leveraging quantum resources to efficiently learn certain properties from nature is a central frontier in quantum information science~\cite{aharonovQuantum2022, bubeckEntanglement2020, chenThe2023, chenExponential2022, chenWhen2023, huangInformation-Theoretic2021, huangQuantum2022}. Both theory and experiment have demonstrated that quantum computers can exhibit significant advantages over classical methods in specific computational problems and various learning tasks~\cite{shorAlgorithms1994, aaronsonThe2011, harrowQuantum2017, aruteQuantum2019, zhongQuantum2020, aharonovQuantum2022, huangInformation-Theoretic2021}. However, these powerful quantum devices are also susceptible to noise. Accurately characterizing quantum noise is a cornerstone for developing fault-tolerant quantum computers~\cite{shorFault-Tolerant1996, gottesmanTheory1998, singhMid-Circuit2023}. The Pauli channel is a noise model of crucial importance in both theory and practice. Under realistic assumptions, many noise types can be transformed into corresponding Pauli channels through techniques such as randomized compiling~\cite{wallmanNoise2016, hashimRandomized2021}. Consequently, efficiently learning an unknown Pauli channel is essential for numerous applications, including quantum benchmarking~\cite{erhardCharacterizing2019, harperEfficient2020, carignan-dugasThe2023}, noise mitigation~\cite{vandenbergProbabilistic2023, ferracinEfficiently2024, kimEvidence2023}, and quantum error correction~\cite{tuckettUltrahigh2018}.

Recently, exploring how quantum resources can gain advantages in learning Pauli channels has garnered attention~\cite{fujiwaraQuantum2003, chiuriExperimental2011, flammiaEfficient2020, flammiaPauli2021, harperFast2021, chenQuantum2022, chenLearnability2023, chenTight2024, fawziLower2025, chenEfficient2025}. One class of schemes, known as concatenated protocols, allows for multiple channel queries and data post-processing within a single measurement round. The advantages of concatenated protocols in learning Pauli channels have been deeply investigated in Refs.~\cite{chenQuantum2022, chenTight2024, fawziLower2025, chenEfficient2025}. In particular, Ref.~\cite{chenEfficient2025} combines concatenated protocols with quantum memory to achieve an exponential advantage in measurement complexity. To learn an $n$-qubit Pauli channel, their protocol requires sequentially testing each of the $4^n - 1$ non-trivial eigenvalues and recording the results into quantum memory before a final measurement is performed. For the outcome to be sufficiently reliable, the error probability of a single test must be suppressed to be exponentially small. To this end, the protocol in Ref.~\cite{chenEfficient2025} utilizes $O(\log n)$ ancilla qubits as independent samples, employing a statistical counting algorithm to amplify the difference between competing hypotheses. In the era of Noisy Intermediate-Scale Quantum (NISQ) computing, high-quality qubits are a precious resource~\cite{lvovskyOptical2009, hetetMultimodal2008, reimSingle-Photon-Level2011, preskillQuantum2018, yuInterferometry-Integrated2023, thomasDeterministic2024}. This context directly leads to a key open question: can we reduce the quantum memory requirement to, for instance, $O(1)$, while preserving the exponential advantage in measurement complexity?

In this work, we provide an affirmative answer to this question by designing a quantum algorithm based on a ``weakly-driven quantum walk.'' The core idea is to leverage the ability of concatenated protocols to query the channel multiple times to drive a ``pointer'' system in a quantum walk. By exploiting its distinct dynamical properties under biased versus unbiased driving, we achieve an exponential suppression of the test error probability. The eigenvalues of a Pauli channel are real-valued parameters, and their measurement requires probabilistic sampling, rendering the driving information classical and random. In the presence of small eigenvalues, driving the pointer with classical probabilistic information carried by this high-von-Neumann-entropy mixed state would cause the pointer to undergo decoherence, eliminating quantum superpositions between different evolutionary paths. This prevents our algorithm from leveraging the superposition of paths for an advantage, as is done in traditional quantum walks~\cite{aharonovQuantum1993, aharonovQuantum2001, childsExponential2003, lyuLocalization2015, gongQuantum2021, wangImplementation2022}. Our approach also differs fundamentally from open quantum walks, which are typically driven by dissipative environmental interactions rather than by classical information generated by a quantum channel~\cite{attalOpen2012, sinayskiyOpen2019}. It should also be distinguished from driven quantum walks, where an external field dynamically creates and annihilates walkers via a nonlinear process, rather than guiding the evolution of a single walker with classical information~\cite{hamiltonDriven2014, heldDriven2022}.

To attain a quantum advantage under such conditions, we devise the ``weak driving'' model. Its central concept is similar to that of weak measurement~\cite{menskyQuantum1979, aharonovHow1988, belavkinQuantum1992, hostenObservation2008, wuWeak2011, pangWeak2012, wuState2013}: weak measurements introduce a weak interaction between the system to be measured and a pointer to minimally disturb the system's coherence. Analogously, our weakly-driven quantum walk model sets each step of the drive to be sufficiently weak, such that even when driven by classical probabilistic information, the pointer system's own quantum coherence is maximally preserved over a long-range evolution. This enables our scheme to utilize the advantage of superposition within the continuous state space of a single qubit's Bloch sphere. By accumulating information into the evolution of the pointer's state, we circumvent the limitations imposed by performing collective operations on multiple, independently prepared samples to gather statistics, a process inherently bound by discrete sampling.

By incorporating our algorithm into the protocol framework of Ref.~\cite{chenEfficient2025}, we can reduce the required quantum memory from $O(\log n)$ to $O(1)$ while preserving its poly($n$) advantage in measurement complexity. Furthermore, we briefly discuss how leveraging weak driving dynamics may enable attaining quantum advantages in other similar quantum algorithms, potentially aiding quantum sensing and precision measurement in resource-constrained scenarios.

\section{Problem Description and Computational Resources}
\label{sec:problem_description}

The main result of this work is centered on a quantum hypothesis testing framework. We consider two mutually exclusive hypotheses regarding the properties of an unknown single-qubit diagonal input state $\rho_{\text{in}}$. For simplicity, we assume these two hypotheses are exhaustive:
\begin{itemize}
    \item \textbf{Null hypothesis ($H_0$):} The input state $\rho_{\text{in}}$ is the maximally mixed state, $I/2$.
    \item \textbf{Alternative hypothesis ($H_1$):} The input state is $\rho_{\text{in}} = (1/2 + \varepsilon^*)\ket{0}\bra{0} + (1/2 - \varepsilon^*)\ket{1}\bra{1}$. Here, $\varepsilon^*$ denotes the true polarization strength of the input state and is a real number satisfying $\varepsilon \le |\varepsilon^*| \le 1/2$, where $\varepsilon$ is a prespecified minimum threshold for our algorithm, with $0 \le \varepsilon \le 1/2$.
\end{itemize}

Our goal is to design a quantum algorithm with no intermediate measurements. For a given target distinguishability parameter $\gamma$, the algorithm produces a final state on a recorder memory qubit, $M$. A single measurement on $M$ is then performed to distinguish between the two hypotheses, such that at least one of the error probabilities is suppressed to $\delta = O(e^{-\gamma})$. Here, $\delta$ is either the Type I error probability (falsely rejecting $H_0$) or the Type II error probability (falsely accepting $H_0$). It is worth noting that while this construction differs from the typical goal of simultaneously suppressing both error probabilities, we will later show that it is sufficient for the task of Pauli channel estimation.

To achieve this, we assume access to the following idealized computational resources:
\begin{enumerate}
    \item \textbf{Constant quantum memory:} The algorithm can use $O(1)$ noiseless and decoherence-free ancilla qubits, which serve as quantum memory to store intermediate information or as working qubits for computation.
    \item \textbf{Universal quantum gates:} The algorithm can employ universal and error-free quantum gates in its circuit.
    \item \textbf{Qubit reset capability:} The algorithm can deterministically reset a specified qubit to the $\ket{0}$ state within the circuit via a measurement-free quantum channel.
    \item \textbf{Concatenated channel queries:} The algorithm can serially query a quantum channel that outputs the state $\rho_{\text{in}}(\varepsilon^*)$ multiple times, with each output being independent of the others.
\end{enumerate}

\section{Design Philosophy and Algorithmic Components}
\label{sec:design_philosophy}

\subsection{Strongly-Driven and Weakly-Driven Quantum Walks}
\label{subsec:driven_walks}

Given our capability for concatenated channel queries, we can subject a quantum system to a dynamical process driven by the signal from the input state. By leveraging the distinct dynamical properties of this process under biased versus unbiased driving, we can create a more pronounced difference in the output state corresponding to the two hypotheses. The quantum walk model we choose implements this dynamical process by repeatedly applying a single-step evolution channel, $\mathcal{C}_{\text{walk}}(\rho_{\text{in}} \otimes \rho_P)$, to a ``pointer'' quantum system $P$. The channel is driven by the input state $\rho_{\text{in}}(\varepsilon^*)$, which is independently prepared in each step. Under unbiased driving ($H_0$), this model exhibits purely diffusive dynamics, whereas under biased driving ($H_1$), it presents a drift-diffusion dynamics.

We now define two distinct types of quantum walks:

\begin{definition}[Strongly-Driven Quantum Walk]
\label{def:strong_drive}
Let the Hilbert space of the pointer system, $\mathcal{H}_P$, be spanned by a complete orthonormal basis $\{\ket{j}_P\}$, and let the Hilbert space of the input system be $\mathcal{H}_{\text{in}}$. The single-step evolution channel of the quantum walk, $\mathcal{C}_{\text{walk}}^{\text{s}}$, is realized by a unitary operator $U_{\text{walk}}^{\text{s}}$ acting on the composite space $\mathcal{H}_{\text{in}} \otimes \mathcal{H}_P$. The walk is defined as strongly-driven if and only if for any basis state $\ket{i}_P \in \{\ket{j}_P\}$ and any input state $\rho_{\text{in}}$,
\begin{equation}
    \bra{i}_P \text{Tr}_{\text{in}}\left(U_{\text{walk}}^{\text{s}}\left(\rho_{\text{in}} \otimes \ket{i}_P\bra{i}_P\right)U_{\text{walk}}^{\text{s}\dagger}\right) \ket{i}_P = 0,
\end{equation}
where $\text{Tr}_{\text{in}}(\cdot)$ denotes the partial trace over the input system.
\end{definition}

This condition implies that after a single step of evolution starting from any basis state $\ket{i}_P$, the probability of finding the pointer system back in that same state is zero. This causes the walk to manifest as a ``jump'' between different orthonormal basis states.

In contrast, we define the weakly-driven quantum walk as follows:

\begin{definition}[Weakly-Driven Quantum Walk]
\label{def:weak_drive}
In a pointer Hilbert space spanned by a complete orthonormal basis $\{\ket{j}_P\}$, a quantum walk with a single-step evolution channel $\mathcal{C}_{\text{walk}}^{\text{w}}$ is defined as weakly-driven if and only if for any basis state $\ket{i}_P \in \{\ket{j}_P\}$ and any input state $\rho_{\text{in}}$, its single-step evolution satisfies:
\begin{equation}
    \bra{i}_P \text{Tr}_{\text{in}}\left( U_{\text{walk}}^{\text{w}} \left( \rho_{\text{in}} \otimes \ket{i}_P\bra{i}_P \right) U_{\text{walk}}^{\text{w}\dagger} \right) \ket{i}_P = 1 - \eta,
\end{equation}
where $\eta$ is a small positive number related to the driving strength.
\end{definition}

This condition indicates that after a single step of evolution starting from any basis state $\ket{i}_P$, the resulting state of the pointer system has a fidelity $\sqrt{1 - \eta}$ close to 1 with respect to the initial state.

In the algorithm proposed in this work, the pointer is a single qubit. We let it perform a one-dimensional walk within the XZ-plane of the Bloch sphere through rotations about the Y-axis. Its unitary operator $U_{\text{walk}}^{\text{w}}$ is given by:
\begin{equation}
    U_{\text{walk}}^{\text{w}} = \ket{0}\bra{0}_{\text{in}} \otimes R_y(+\theta)_P + \ket{1}\bra{1}_{\text{in}} \otimes R_y(-\theta)_P,
\end{equation}
where $R_y(\pm\theta)_P$ are rotation operators acting on the pointer qubit $P$, corresponding to a rotation by an angle $\pm\theta$ about the Y-axis of the Bloch sphere. The angle $\theta$ is a small quantity determined by the target parameter $\gamma$ and the threshold $\varepsilon$.

Suppose that at the beginning of a step, the pointer $P$ is in the state $\rho_P$ and the input is $\rho_{\text{in}} = (1/2 + \varepsilon^*)\ket{0}\bra{0} + (1/2 - \varepsilon^*)\ket{1}\bra{1}$. After one walk operation, the reduced density matrix of the pointer qubit becomes $\rho'_P = (1/2 + \varepsilon^*) \cdot \mathcal{E}_{+\theta}(\rho_P) + (1/2 - \varepsilon^*) \cdot \mathcal{E}_{-\theta}(\rho_P)$, where $\mathcal{E}_{\pm\theta}(\rho) = R_y(\pm\theta)\rho R_y(\pm\theta)^\dagger$. This result shows that because our input state is a diagonal density matrix, the information relevant to our hypothesis test exists only in the form of classical probabilities. Consequently, the new state $\rho'_P$ after one step of evolution is a classical probabilistic mixture of two different evolutionary paths (a $+\theta$ rotation and a $-\theta$ rotation). This is in stark contrast to traditional quantum walk algorithms, which leverage quantum coherence between different paths to gain an advantage~\cite{aharonovQuantum1993, aharonovQuantum2001, childsExponential2003, lyuLocalization2015, gongQuantum2021, wangImplementation2022}. 

To illustrate the meaning and advantage of weak driving, we start with the pointer in the initial state $\ket{0}_P\bra{0}_P$ and calculate the purity of its state after one step, which yields $\text{Tr}[(\rho'_P)^2] = 1 - [(1 - 4\varepsilon^{*2})\sin^2\theta] / 2$. For a small step size $\theta$, we have $\sin\theta \approx \theta$, and thus the single-step purity loss is $\Delta P = 1 - \text{Tr}[(\rho'_P)^2] \approx (1 - 4\varepsilon^{*2})\theta^2 / 2 \le \theta^2 / 2$.

This result demonstrates that even when driven by classical probabilistic information, the pointer system does not suffer from significant decoherence after a single step. This idea is similar to the paradigm of weak measurement: weak measurements introduce a weak interaction between the system to be measured and a pointer, which, even after a decohering projective measurement on the pointer, minimally disturbs the system's coherence~\cite{menskyQuantum1979, aharonovHow1988, belavkinQuantum1992, hostenObservation2008, wuWeak2011, pangWeak2012, wuState2013}. Our weakly-driven quantum walk introduces a weak driving interaction on the input and pointer systems. Consequently, even if the input information is a classical probability, it does not cause significant decoherence in the pointer system. It is precisely this weak driving that allows the single-qubit pointer state to maintain its coherence during a long-range walk. This enables us to leverage the advantage of superposition in the continuous state space of a single qubit's Bloch sphere, thereby circumventing the limitations of discrete sampling associated with accumulating statistical information via collective operations on multiple, independently prepared samples.

\subsection{Serial Overwriting and Qubit Reset}
\label{subsec:writing_reset}

The trade-off of the weak driving approach is twofold. First, the high overlap between pointer states after consecutive steps limits the ability to distinguish the two hypotheses with a single measurement. Second, for any control operation performed by the pointer, tracing out the controlled system leads to the decoherence of the pointer itself, rendering it unusable for subsequent steps of the walk.

To address the first issue, we consider a ``controlled-overwrite channel,'' $\mathcal{C}_{\text{write}}$. After each round of the walk, we perform an irreversible write operation on the recorder memory qubit $M$, conditioned on the state of the pointer $P$. This channel, $\mathcal{C}_{\text{write}}$, corresponds to a complete set of Kraus operators:
\begin{align*}
    K_0 &= \ket{0}\bra{0}_P \otimes I_{M}, \\
    K_{1,0} &= \ket{1}\bra{1}_P \otimes \ket{1}\bra{0}_{M}, \\
    K_{1,1} &= \ket{1}\bra{1}_P \otimes \ket{1}\bra{1}_{M}.
\end{align*}
The function of this channel is to deterministically set the state of the recorder qubit $M$ to $\ket{1}$ if the pointer $P$ is in the $\ket{1}$ state; otherwise, if $P$ is in the $\ket{0}$ state, the state of $M$ remains unchanged. Notably, this channel induces decoherence of the pointer into the computational basis.

Suppose that before a controlled-overwrite operation, the pointer $P$ is in state $\rho_P$, and the recorder qubit $M$ is in a diagonal mixed state $\rho_M = p\ket{0}\bra{0} + (1-p)\ket{1}\bra{1}$, where $p$ is the probability of $M$ being in the $\ket{0}$ state. After the write operation $\mathcal{C}_{\text{write}}$, the probability that $M$ remains in the $\ket{0}$ state is given by $p' = \bra{0}\left( \text{Tr}_P(\mathcal{C}_{\text{write}}(\rho_P \otimes \rho_M)) \right)\ket{0} = \bra{0}\rho_P\ket{0} \cdot p$. This recurrence relation shows that after each write operation, the probability of $M$ remaining in the $\ket{0}$ state is the product of its previous probability and the probability of finding the pointer $P$ in the $\ket{0}$ state during that operation. We term the probability of $M$ remaining in the $\ket{0}$ state the ``survival probability''. It is by leveraging the multiplicative nature of this survival probability under serial controlled overwriting that we can distinguish the two hypotheses with an extremely low error probability after a final measurement on $M$.

Since our algorithm features no intermediate measurements, this channel must be constructed via Stinespring dilation. This requires the dimension of the ancilla space to be at least equal to the number of Kraus operators, which can be realized by introducing two ancilla qubits to provide a $4$-dimensional ancilla space. The serial nature of our algorithm ensures that these two ancillas can be reset and reused between overwrite operations, thus not affecting the constant-order advantage of our algorithm in terms of ancilla qubit resources.

Beyond the decoherence caused by using the pointer for control operations, a more fundamental challenge lies in the fact that the core dynamics of the walk are driven by a mixed state $\rho_{\text{in}}$ with high von Neumann entropy. As indicated by our earlier purity loss calculation, although the weak-driving design minimizes the entropy influx from each controlled operation, entropy still serially flows from independent copies of $\rho_{\text{in}}$ and accumulates in the pointer $P$. This leads to a continuous degradation of the pointer's coherence, fundamentally inhibiting our ability to distinguish the two hypotheses. Therefore, it is necessary to perform a non-unitary reset process before the start of each round in the serial algorithm, to actively ``pump out'' the accumulated entropy from the pointer to an external environment.

Fortunately, well-developed schemes exist for the measurement-free cooling and reset of qubits~\cite{boykinAlgorithmic2002, geerlingsDemonstrating2013, magnardFast2018, alhambraHeat-Bath2019, aamirThermally2025}. A direct implementation involves introducing an ancilla qubit $A$ that can be coupled to an external heat bath and efficiently cooled to its ground state $\ket{0}_A$. The heat bath is external to the algorithm, and its cooling timescale is much shorter than an algorithmic cycle. When the algorithm needs to reset a qubit, a SWAP gate is applied between it and the ancilla $A$, exchanging their states. Subsequently, the entropy-carrying ancilla $A$ interacts with the heat bath and is re-cooled to $\ket{0}_A$, preparing it for the next reset operation. Again, the serial nature of our algorithm allows this ancilla to be reused, preserving our advantage in requiring only a constant number of ancilla qubits.

\subsection{Multi-Round Independent Walks with Varying Steps}
\label{subsec:multi_round_walks}

A rotation-based walk within a finite state space, such as the pointer's Bloch sphere, faces two issues if a fixed step count is used across multiple independent rounds: periodicity and non-monotonicity. Periodicity implies that for two total rotation angles, $\Theta$ and $\Theta'$, any measurement will fail to distinguish between them if their difference is an integer multiple of $2\pi$. Associated with periodicity is the non-monotonicity of the protocol's response. While we expect the algorithm to satisfy the condition that a stronger signal yields a more pronounced output response, a challenge arises. If we choose the walk parameters (e.g., total steps $i$ and single-step angle $\theta$) to be highly sensitive to a small $|\varepsilon^*|$, the outcome for a larger $|\varepsilon^*|$ might fall into a low-response region of a different periodic cycle.

Considering that our algorithm possesses the capabilities of serial overwriting and measurement-free reset, we introduce a ``multi-round independent walks with varying steps'' mechanism to address these two problems. Specifically, we perform $m$ independent rounds, with the $i$-th round using exactly $i$ steps (for $i = 1, 2, \dots, m$). For appropriately chosen parameters $m$ and $\theta$, this mechanism ensures that for a small number of steps, the algorithm's response increases monotonically with the signal strength $|\varepsilon^*|$. Even as the response in individual rounds becomes periodic with an increasing number of steps, the serial overwriting mechanism accumulates the probability of $M$ remaining in the $\ket{0}$ state. This ensures that outcomes from high-response regions are effectively preserved, thereby creating a significant and detectable difference between the null and alternative hypotheses.

\section{Algorithm and Performance Analysis}
\label{sec:algorithm_analysis}

Integrating the design philosophy and algorithmic components, we present our algorithm based on the weakly-driven quantum walk in Algorithm~\ref{alg:single_stage_algorithm}.

\begin{algorithm}[H]
\caption{Hypothesis Testing via a Weakly-Driven Quantum Walk}
\label{alg:single_stage_algorithm}
\begin{algorithmic}[1]
\Resources \parbox[t]{0.85\linewidth}{
    Access to the channel that prepares $\rho_{\text{in}}(\varepsilon^*)$ \\
    A sample qubit $Q_S$ \\
    A pointer qubit $P$ \\
    A recorder memory qubit $M$
}
\Statex
\Require Total number of rounds $m$; single-step angle $\theta$.
\Ensure The final state of the recorder memory qubit $M$.
\Statex
\State \textbf{Initialization:} $P \to \ket{0}\bra{0}$, $M \to \ket{0}\bra{0}$.
\For{round $i = 1, \dots, m$}
    \State $P \to \ket{0}\bra{0}$.
    \For{step $j = 1, \dots, i$}
        \State Prepare $\rho_{\text{in}}(\varepsilon^*)$ on $Q_S$.
        \State Apply $\mathcal{C}_{\text{walk}}^{\text{w}}$ on $Q_S$ and $P$.
    \EndFor
    \State Apply $\mathcal{C}_{\text{write}}$ on $P$ and $M$.
\EndFor
\end{algorithmic}
\end{algorithm}

After the algorithm concludes, we perform a single measurement on the output recorder qubit $M$ in the computational basis and make a judgment on the two hypotheses based on the outcome. This requires calculating the probability of obtaining the outcome $\ket{0}$ under both the $H_0$ and $H_1$ hypotheses. This probability, which we previously defined as the ``survival probability,'' is a function of the number of rounds $m$, the angle $\theta$, and the signal strength $\varepsilon^*$, denoted as $S(m, \theta, \varepsilon^*)$.

By approximating the discrete binomial distribution with a Gaussian distribution, performing Taylor expansions on the relevant functions and retaining only their leading-order terms, we can derive an approximate logarithmic expression for the survival probability:
\begin{equation}
    -\ln S(m, \theta, \varepsilon^*) \approx  \frac{m^2\theta^2}{8} + \frac{m^3\varepsilon^{*2}\theta^2}{3}.
    \label{eq:log_survival}
\end{equation}
The detailed derivation of this formula and an analysis for the validity of the approximation are presented in Appendix~\ref{app:calc_and_approx}. This result reveals a background decay term, $m^2\theta^2/8$, arising from pure diffusion (which is the only term present under $H_0$), and a signal-induced decay term, $m^3\varepsilon^{*2}\theta^2/3$, arising from the biased drift. Crucially, in this logarithmic expression, the background decay is a quadratic function of the number of rounds $m$, whereas the signal-induced decay is a cubic function. Therefore, by choosing appropriate parameters, we can create a sufficiently large difference between the survival probabilities under the two hypotheses.

We note that the survival probability under $H_1$ is always less than that under $H_0$. A natural decision rule is as follows: if the measurement outcome is $\ket{0}$, we accept $H_0$; if it is $\ket{1}$, we reject $H_0$. Under this rule, the Type I error probability is $\alpha = 1 - S_0$, and the Type II error probability is $\beta = S_1$, where $S_0$ and $S_1$ are the survival probabilities of the recorder qubit under the null and alternative hypotheses, respectively.

Considering the worst-case scenario where $|\varepsilon^*| = \varepsilon$, our goal is to suppress at least one type of error probability to $O(e^{-\gamma})$. We choose to target the Type II error, $\beta$. Additionally, for an effective hypothesis testing protocol, we require the Type I error, $\alpha$, to be bounded by a constant. This leads to the following set of constraints:
\begin{align}
    -\ln S_0 &\approx \frac{m^2\theta^2}{8} = O(1) \implies m\theta = O(1), \\
    \ln(S_0/S_1) &= \ln S_0 - \ln S_1 \approx \frac{m^3\varepsilon^2\theta^2}{3} = \Omega(\gamma).
\end{align}
Solving these constraints yields $m = \Omega(\gamma / \varepsilon^2)$ and $\theta = O(\varepsilon^2 / \gamma)$. With such a choice of parameters, our algorithm can suppress the Type II error probability $\beta$ to a desired exponential scale, $e^{-\gamma}$, while controlling the Type I error probability $\alpha$ at a constant level.

The approximations we employ include approximating the discrete binomial distribution with a continuous Gaussian distribution, applying a continuous approximation to the summation formulas, and performing Taylor expansions on the trigonometric, exponential, and logarithmic functions while retaining only the leading-order terms. The errors introduced by these approximations become significant under the following conditions:  (i) when the single-step rotation angle $\theta$ is large (as the contribution from higher-order terms in the Taylor expansion increases), and (ii) when the total number of rounds $m$ is small (as the continuous approximation deviates from the discrete summation). This corresponds to scenarios with a small target distinguishability parameter $\gamma$ and a large algorithmic threshold $\varepsilon$.

To visually illustrate the performance of our algorithm and the validity of our approximation under stringent conditions, we choose $\gamma=3.0$ and $\varepsilon=0.25$ for a numerical simulation. Under these parameters, we require the protocol to suppress the Type II error probability to $\beta = S_1 < e^{-\gamma} \approx 0.0498$, while ensuring the Type I error probability satisfies $\alpha = 1 - S_0 < 1/2$. To ensure these targets are met, we introduce a small amount of redundancy in our parameter selection.

\begin{figure}[t]
    \centering
    \includegraphics[width=\columnwidth]{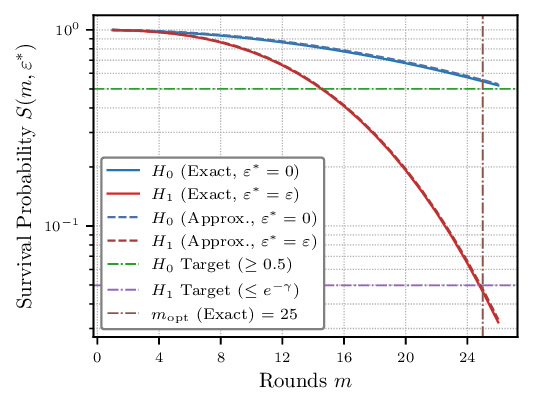}
    \caption{Numerical simulation of the recorder qubit survival probability $S(m, \varepsilon^*)$ as a function of the number of rounds $m$, for $\gamma=3.0$ and $\varepsilon=0.25$. The solid lines represent the exact numerical results based on the binomial distribution, while the dashed lines show the predictions from our analytical formula, Eq.~(\ref{eq:log_survival}). The blue curves correspond to the null hypothesis $H_0$ ($\varepsilon^*=0$), and the red curves correspond to the alternative hypothesis $H_1$ ($\varepsilon^*=\varepsilon$). The plot clearly shows that even in this parameter regime where approximation errors are most pronounced, our analytical formula excellently predicts the system's true dynamical behavior. The green and purple dashed-dotted lines indicate our target survival probability thresholds for $H_0$ and $H_1$, respectively. The brown dashed-dotted line marks the optimal number of rounds, $m_{\text{opt}}=25$, where the protocol simultaneously satisfies both conditions according to the exact calculation, yielding $S_0 \approx 0.55 > 0.5$ and $S_1 \approx 0.0475 < e^{-3}$.}
    \label{fig:validation}
\end{figure}

In Fig.~\ref{fig:validation}, we present the evolution of the survival probability under these parameter settings, comparing the exact numerical calculation based on the binomial distribution and our approximate formula. Consistent with the figure, even in this parameter regime where the approximation errors are most significant, our analytical approximation (dashed lines) remains in excellent agreement with the exact numerical results (solid lines). For the exact calculation, at round $m=25$, the protocol meets the specified performance requirements: $S_0 \approx 0.55 > 0.5$ and $S_1 \approx 0.0475 < e^{-3}$. This visually demonstrates how our scheme creates a distinguishable probabilistic difference by leveraging the different decay dynamics under the two hypotheses.

\section{Learning Pauli Channels}
\label{sec:apply_pauli}

To apply our algorithm to reduce the memory requirements in an $n$-qubit Pauli channel estimation protocol, we do not need to create a new protocol framework. Instead, within the framework proposed in Ref.~\cite{chenEfficient2025}, we replace their \textit{counting scheme}, a subroutine that uses $O(\log(n/\varepsilon^2))$ quantum memory as independent samples to amplify the statistical difference between hypotheses.

First, this requires a method to encode the difference between an unknown non-trivial eigenvalue $\lambda_t$ of an $n$-qubit Pauli channel and its hypothesized value $\hat{\lambda}_t$ (where $t \in \{1, \dots, 4^n-1\}$ and $\lambda_t, \hat{\lambda}_t \in [-1, 1]$) onto a sample ancilla qubit $Q_S$ in the form of our required input state, $\rho_{\text{in}} = (1/2+\varepsilon^*)\ket{0}\bra{0} + (1/2-\varepsilon^*)\ket{1}\bra{1}$. In Appendix~\ref{subapp:deviation_encoding}, we construct a measurement-free quantum circuit to achieve this encoding process. The core idea is as follows: in the framework of Ref.~\cite{chenEfficient2025}, to probe information about the $t$-th Pauli operator $P_t$, one prepares $n$ working qubits in the state $\rho = (I + P_t)/2^n$ and passes them through the unknown channel, resulting in $\rho' = (I + \lambda_t P_t)/2^n$. In the context of Pauli channel estimation, these $n$ working qubits are not counted towards the ancilla overhead, as they are problem-inherent resources for estimating any $n$-qubit Pauli channel. Subsequently, we design a positive operator-valued measure (POVM) that is calibrated and parameterized by the hypothesized value $\hat{\lambda}_t$. With two reusable additional ancilla qubits, we implement this POVM as a measurement-free quantum channel via Stinespring dilation. This channel acts on the output state $\rho'$ and a sample qubit $Q_S$, preparing the reduced density matrix of $Q_S$ into our desired input state, where
\begin{equation}
    \varepsilon^* = \frac{\lambda_t - \hat{\lambda}_t}{2(1 + |\hat{\lambda}_t|)} \in [-\frac{1}{2}, \frac{1}{2}].
\end{equation}
Since $2(1 + |\hat{\lambda}_t|) \le 4$, if the final protocol requires that the difference between all hypothesized and true values, $|\lambda_t - \hat{\lambda}_t|$, is less than an overall precision threshold $\varepsilon_p$, we only need to set the threshold $\varepsilon$ of our subroutine to $\varepsilon = \varepsilon_p / 4 = \Theta(\varepsilon_p)$.

Furthermore, when the hypothesized values are very close to the true values, the outcomes of the hypothesis tests will be dominated by $H_0$ results. If the algorithm's survival probability under $H_0$ is large, the overwhelming $H_0$ outcomes from independent tests of the $4^n-1$ eigenvalues will drown out the $H_1$ signals corresponding to the hypothesized values that need refinement. The \textit{counting scheme} in Ref.~\cite{chenEfficient2025}, after being serially invoked $3n$ times, yields a final survival probability $S_0 = 1/8^n$ and $S_1 \ge e^{-3} = O(1)$. This outcome is the exact opposite of the outcome produced by our Algorithm~\ref{alg:single_stage_algorithm}.

To solve this problem, we design an algorithm with a double-loop structure based on Algorithm~\ref{alg:single_stage_algorithm}, drawing inspiration from Ref.~\cite{chenEfficient2025}. The full details of this ``double-stage'' algorithm are provided in Appendix~\ref{subapp:double-stage}. We denote the recorder qubit in Algorithm~\ref{alg:single_stage_algorithm} as $M_1$ and introduce a second recorder qubit $M_2$. In the inner loop, we implement a subroutine based on Algorithm~\ref{alg:single_stage_algorithm}, choosing $m=O(\log n / \varepsilon^2)$ and $\theta=O(\varepsilon^2 / \log n)$; this subroutine outputs an $M_1$ survival probability that satisfies $S_0^{(1)} \ge 1/2$ and $S_1^{(1)} \le 1/n$. We then use $M_1$ as the control qubit and $M_2$ as the target qubit, applying a ``reverse controlled-overwrite channel'' $\mathcal{C}_{\text{r-write}}$. The term ``reverse'' indicates that its logic is opposite to that of the controlled-overwrite channel $\mathcal{C}_{\text{write}}$: if $M_1$ is in the $\ket{0}$ state, $M_2$ is overwritten to $\ket{1}$; otherwise, $M_2$'s state is unchanged.

Under this operation, $M_2$'s single-iteration survival probability is $1 - S_0^{(1)} \le 1/2$ under $H_0$, and $1 - S_1^{(1)} \ge 1 - 1/n$ under $H_1$. In the outer loop, to match the architecture of the protocol in Ref.~\cite{chenEfficient2025}, this process is serially repeated $3n$ times. Now, with $M_2$ as the final output, its overall survival probability under $H_0$ is $S_0^{(2)} \le (1/2)^{3n} = 1/8^n$, while its overall survival probability under $H_1$ is $S_1^{(2)} \ge (1-1/n)^{3n} \approx e^{-3}$.

In Sec.~\ref{sec:problem_description}, for model simplicity, we assumed the two hypotheses our algorithm distinguishes are exhaustive, and did not consider true signals with $0 < |\varepsilon^*| < \varepsilon$. In the protocol of Ref.~\cite{chenEfficient2025}, when testing the $4^n-1$ eigenvalues, the output of each test (which involves invoking their \textit{counting scheme} $3n$ times) is not directly measured. Instead, it is used as a control qubit to record its information via a controlled channel onto a final recorder qubit, $M_3$, which is not reset throughout the $4^n-1$ tests. In a realistic scenario, many non-zero but sub-threshold true signals could interfere with the record on $M_3$. This necessitates an analysis of our algorithm's performance under such conditions. In Appendix~\ref{subapp:pauli_performance}, we demonstrate that in the actual task of Pauli channel estimation, our algorithm's performance meets expectations and aligns with the requirements of the protocol in Ref.~\cite{chenEfficient2025}.

In summary, by integrating our algorithm into the protocol framework of Ref.~\cite{chenEfficient2025}, we can accomplish the task of estimating the eigenvalues of an unknown $n$-qubit Pauli channel. Our algorithm requires no intermediate measurements, with all parameters specified upfront. Therefore, we reduce the quantum memory requirement from logarithmic to constant order, while preserving the non-adaptive nature and the advantage in measurement complexity of the protocol in Ref.~\cite{chenEfficient2025}. In terms of channel query complexity, a single run of our double-stage subroutine requires $3n \times (\sum_{i=1}^{m} i) = O(n \cdot m^2) = O(n \log^2 n / \varepsilon^4)$ queries. This is an increase from the $O(n \log n / \varepsilon^2)$ queries required for $3n$ serial invocations of the \textit{counting scheme} in the original work. This can be viewed as a trade-off for the simple varying-step design used to overcome periodicity. We note, however, that this quadratic scaling in $m$ is not fundamental to the weak-driving approach itself. A more optimized protocol could employ a sparse set of $O(1)$ large, co-prime step numbers, each on the order of $m$, to resolve periodicity. Such a scheme would achieve a total query complexity comparable to that of Ref.~\cite{chenEfficient2025}, while retaining the constant memory advantage.

\section{Discussion and Outlook}
\label{sec:outlook}

It is proved in Ref.~\cite{chenEfficient2025} that for any concatenated protocol using $k$ ancilla qubits as memory to estimate Pauli channel eigenvalues with constant error, at least $\Omega(2^{(n-k)/3})$ channel queries are required. This implies that a Pauli channel estimation protocol using constant-order quantum memory and poly($n$) measurements would require the memory qubits to remain coherent over an exponential number of queries, imposing a stringent requirement on the memory's coherence time. Thus, a trade-off critical for NISQ-era quantum computers emerges among measurement complexity, quantum memory qubit count, and memory performance. Given our algorithm's advantage in quantum memory count, further exploring its utility within this trade-off highlights a research direction critical for implementing such Pauli channel learning algorithms on near-term quantum platforms.

Meanwhile, our core mechanism shows promise for applications in learning tasks for other quantum states and processes, especially in scenarios involving decoherence and non-unitary processes. From a broader perspective, incorporating dynamical processes akin to quantum walks into quantum algorithms and leveraging the system's dynamical response characteristics to gain a quantum advantage (as exemplified by our ``weak driving'' mechanism), while avoiding excessive additional resource requirements, could be a valuable strategy. This approach could facilitate designing similar quantum algorithms and advance quantum sensing and precision measurement of weak, noisy signals in resource-constrained scenarios.

\begin{acknowledgments}
This work is supported by the National Natural Science Foundation of China (Grants 12175104 and 12475020), the Innovation Program for Quantum Science and Technology (2021ZD0301701), and the National Key Research and Development Program of China (2023YFC2205802).
\end{acknowledgments}

\appendix

\section{Derivation of the Survival Probability and Validity of the Approximations}
\label{app:calc_and_approx}

This appendix provides a mathematical derivation for the core formula, Eq.~(\ref{eq:log_survival}), and justifies the validity of the approximations.

\subsection{Derivation of the Survival Probability}

In our algorithm, the $i$-th round (for $i \in \{1, \dots, m\}$) consists of an $i$-step quantum walk driven by the input state $\rho_{\text{in}}(\varepsilon^*)$. Let $n_+$ be the number of steps with a $+\theta$ rotation and $n_-$ be the number of steps with a $-\theta$ rotation, such that the total number of steps is $i = n_+ + n_-$. The final total rotation angle for the pointer is $\Theta_i = (n_+ - n_-)\theta$. Since the choice at each step is independent, the number of steps $n_+$ follows a binomial distribution $B(i, p)$, where $p = 1/2+\varepsilon^*$. For a sufficiently large $i$, this discrete binomial distribution can be well approximated by a continuous Gaussian distribution. Consequently, the distribution of the total rotation angle $\Theta_i$ also approximates a Gaussian distribution, $N(\mu_i, \sigma_i^2)$, with mean $\mu_i = E[(n_+ - n_-)\theta] = (i p - i(1-p))\theta = i(2p-1)\theta = 2i\varepsilon^*\theta$ and variance $\sigma_i^2 = \text{Var}((n_+ - n_-)\theta) = \theta^2 \cdot i \cdot 4p(1-p) = i\theta^2(1-4\varepsilon^{*2})$.

After this round, the probability that the recorder qubit is overwritten is equal to the probability of measuring the pointer in the $\ket{1}$ state, $P(P_i=\ket{1}) = E[\sin^2(\Theta_i/2)]$. Using the identity $\sin^2(x) = (1-\cos(2x))/2$, we have:
\begin{equation}
    P(P_i=\ket{1}) = \frac{1}{2} \left(1 - E[\cos(\Theta_i)]\right).
\end{equation}
For a Gaussian random variable $X \sim N(\mu, \sigma^2)$, the expectation of its cosine follows a standard formula: $E[\cos(X)] = e^{-\sigma^2/2}\cos(\mu)$. Substituting our parameters, we obtain the expression under the Gaussian approximation for the single-round overwrite probability:
\begin{equation}
    P(P_i=\ket{1}) = \frac{1}{2} \left(1 - e^{-i\theta^2(1-4\varepsilon^{*2})/2} \cos(2i\varepsilon^*\theta)\right).
\end{equation}
The overall survival probability of the recorder qubit, denoted as a function $S(m, \theta, \varepsilon^*)$, is the product of the survival probabilities from all rounds: $S = \prod_{i=1}^{m} (1 - P(P_i=\ket{1}))$. Its logarithm is $\ln S = \sum_{i=1}^{m} \ln(1 - P(P_i=\ket{1}))$.

Now, we perform a Taylor expansion on the expression and retain up to the leading-order terms. Assuming that both $i\theta^2$ and $\varepsilon^*$ (and thus $i\varepsilon^*\theta$) are small quantities, we get:
\begin{equation}
    \begin{aligned}
        P(P_i=\ket{1}) 
        &\approx \frac{1}{2} \Biggl(1 - \left(1 - \frac{i\theta^2(1-4\varepsilon^{*2})}{2}\right) \\
        &\qquad \times \left(1 - \frac{(2i\varepsilon^*\theta)^2}{2}\right)\Biggr) \\
       &= \frac{i\theta^2}{4} + i^2\varepsilon^{*2}\theta^2 + O(i\varepsilon^{*2}\theta^2,i^3\varepsilon^{*2}\theta^4)
    \end{aligned}
    \label{eq:overwrite_prob_approx}
\end{equation}
Next, we sum this expression to calculate the total logarithmic survival probability. Using the summation formulas $\sum_{i=1}^m i = \frac{m(m+1)}{2} \approx \frac{m^2}{2}$ and $\sum_{i=1}^m i^2 = \frac{m(m+1)(2m+1)}{6} \approx \frac{m^3}{3}$, and neglecting higher-order terms, we obtain:
\begin{equation}
    \begin{aligned}
    -\ln S(m, \theta, \varepsilon^*) &\approx \sum_{i=1}^{m} \left(\frac{i\theta^2}{4} + i^2\varepsilon^{*2}\theta^2\right) \\
    &\approx \frac{m^2\theta^2}{8} + \frac{m^3\varepsilon^{*2}\theta^2}{3}.
    \end{aligned}
\end{equation}

\subsection{Justification for the Validity of the Approximations}

The approximations used to obtain the above results include: approximating the discrete binomial distribution with a continuous Gaussian distribution, applying a continuous approximation to the summation formulas, and performing Taylor expansions on the trigonometric, exponential, and logarithmic functions while retaining only the leading-order terms. In Fig.~\ref{fig:validation}, by comparing with exact numerical calculations, we have already visually demonstrated that our composite approximate model remains highly effective even for a small total number of rounds, $m$.

Concurrently, in our Taylor expansion, we assumed that $i\theta^2$ and $i\varepsilon^*\theta$ are small quantities. The maximum value of $i\theta^2$ is $m\theta^2$. Based on our final parameter choices ($m = \Omega(\gamma / \varepsilon^2)$ and $\theta = O(\varepsilon^2 / \gamma)$), and given the freedom we have in selecting the constant factors, we can ensure that this approximation holds for the vast majority of rounds $i$. However, for the alternative hypothesis $H_1$, when the true signal strength $\varepsilon^*$ is much larger than our specified threshold $\varepsilon$, the drift term $i\varepsilon^*\theta$ may no longer be a small quantity. This would not only affect the approximation of the trigonometric functions but could also cause the higher-order terms we neglected in the logarithmic expression, $O(i\varepsilon^{*2}\theta^2, i^3\varepsilon^{*2}\theta^4)$, to make a non-negligible contribution.

Although our approximation may no longer be valid when the signal strength $\varepsilon^*$ is much larger than the threshold $\varepsilon$, this does not affect the validity of our final conclusion. The reason is that the multi-round, varying-step walk design ensures that the protocol's response is monotonically enhanced with increasing $\varepsilon^*$. The primary purpose of our approximate calculation is to analyze the protocol's performance in the most challenging case, i.e., when the signal strength is near the small threshold, $\varepsilon^*=\varepsilon$. As long as we can demonstrate that the protocol succeeds in this scenario, its performance will only be better for a larger $\varepsilon^*$, making it even more clearly distinguishable from the null hypothesis. Therefore, the parameter choices derived from our small-quantity approximation are both reasonable and robust.

\begin{figure}[t]
    \centering
    \includegraphics[width=\columnwidth]{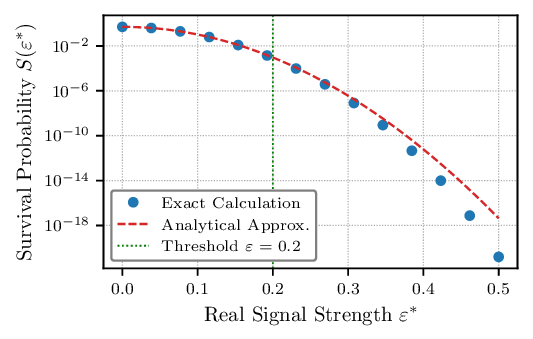}
    \caption{The final survival probability of the recorder qubit, $S(\varepsilon^*)$, as a function of the true signal strength $\varepsilon^*$ for fixed protocol parameters ($m=85, \theta \approx 0.0277$). The blue dots represent the numerical results from an exact calculation based on the binomial distribution, while the red dashed line shows the prediction from our analytical approximation. The vertical green dashed line marks our specified algorithmic threshold of $\varepsilon=0.2$. The plot clearly demonstrates two points: (1) the protocol's response (i.e., the decay) is a monotonic function of the signal strength $\varepsilon^*$; and (2) for small $\varepsilon^*$, our analytical formula is in excellent agreement with the exact calculation.}
    \label{fig:monotonicity}
\end{figure}

To illustrate this point more intuitively, we set the target distinguishability parameter to $\gamma=7.0$ and the algorithmic threshold to $\varepsilon=0.2$. We then calculate the required fixed total number of rounds, $m=85$, and the single-step rotation angle, $\theta \approx 0.0277$. Using both our analytical approximation and exact numerical calculation, we compute the final survival probability of the recorder qubit, $S(\varepsilon^*)$, as a function of varying true signal strength $\varepsilon^*$. The results are shown in Fig.~\ref{fig:monotonicity}.

As shown in Fig.~\ref{fig:monotonicity}, the numerical results clearly prove that the survival probability is a monotonically decreasing function of the true signal strength $\varepsilon^*$. An interesting observation is that for large $\varepsilon^*$, our approximate formula (red dashed line) systematically overestimates the true survival probability (blue dots). This is primarily due to our approximation $\ln(1-x) \approx -x$ (where $x$ is the single-round overwrite probability $P(P_i=\ket{1})$), which significantly underestimates the true amount of decay when $x$ is large. The effect of this error dominates the errors from other approximations. This finding implies that the actual performance of our algorithm is even better than what our approximation predicts, enabling a more effective distinction for stronger signals.

\section{Technical Details for Learning Pauli Channels}
\label{app:learn_pauli}

This appendix details the necessary constructions for applying our algorithm to the $n$-qubit Pauli channel estimation protocol proposed in Ref.~\cite{chenEfficient2025}, and verifies our algorithm's performance within that protocol.

\subsection{Deviation-Encoding Channel}
\label{subapp:deviation_encoding}

To encode the difference between an unknown non-trivial eigenvalue $\lambda_t$ of an $n$-qubit Pauli channel and its hypothesized value $\hat{\lambda}_t$ (where $t \in \{1, \dots, 4^n-1\}$ and $\lambda_t, \hat{\lambda}_t \in [-1, 1]$) onto a sample ancilla qubit in the form of our required input state, $\rho_{\text{in}} = (1/2+\varepsilon^*)\ket{0}\bra{0} + (1/2-\varepsilon^*)\ket{1}\bra{1}$, we first define a parameterized POVM with two outcomes that acts on the $n$ working qubits. Its measurement elements are $\{E_t(\hat{\lambda}_t), I - E_t(\hat{\lambda}_t)\}$.

One of the POVM elements, $E_t(\hat{\lambda}_t)$, is a linear combination of the spectral projectors, $P_{t,\pm} = (I \pm P_t)/2$, of the $t$-th Pauli operator $P_t$:
\begin{equation}
\label{eq:povm_element_def}
E_t(\hat{\lambda}_t) = c_1 P_{t,+} + c_2 P_{t,-},
\end{equation}
where the coefficients $c_1, c_2 \in [0, 1]$ are yet to be determined. When this measurement is applied to the state after it has passed through the unknown channel, $\rho' = \frac{1}{2^n}(I + \lambda_t P_t)$, the probability of obtaining the outcome corresponding to $E_t(\hat{\lambda}_t)$ is:
\begin{equation}
\label{eq:povm_prob}
\begin{aligned}
p(\lambda_t, \hat{\lambda}_t) &= \text{Tr}(E_t(\hat{\lambda}_t) \rho') \\
&= \frac{1}{2} \left[ c_1(1 + \lambda_t) + c_2(1 - \lambda_t) \right].
\end{aligned}
\end{equation}
To achieve the goal of encoding the difference $(\lambda_t - \hat{\lambda}_t)$ into the signal $\varepsilon^*$, the coefficients $c_1$ and $c_2$ must satisfy two conditions:
\begin{enumerate}
    \item \textbf{Zero-deviation calibration:} Under the null hypothesis, i.e., $\lambda_t = \hat{\lambda}_t$, the probability must be exactly $1/2$. This imposes the constraint: $c_1(1 + \hat{\lambda}_t) + c_2(1 - \hat{\lambda}_t) = 1$.
    \item \textbf{Sensitivity maximization:} Subject to the calibration constraint, we must maximize the local response of the probability to changes in $\lambda_t$, which means maximizing the absolute value of its derivative: $|\frac{\partial p}{\partial \lambda_t}| = |\frac{c_1 - c_2}{2}|$.
\end{enumerate}
Solving this constrained optimization problem yields the optimal coefficients. For instance, when $\hat{\lambda}_t \ge 0$, an optimal solution is $c_2=0, c_1 = 1/(1+\hat{\lambda}_t)$. Similarly, when $\hat{\lambda}_t < 0$, an optimal solution is $c_1=0, c_2 = 1/(1-\hat{\lambda}_t)$. Both choices are physically valid.

The next step is to implement this POVM as a measurement-free quantum channel. This process follows the standard Stinespring dilation construction, which embeds a POVM described by a set of measurement operators $\{M_i\}$ into a unitary evolution on an extended Hilbert space.

Taking the case $\hat{\lambda}_t \ge 0$ as an example, the measurement operator corresponding to the POVM element $E_t(\hat{\lambda}_t)$ is $M_1 = \sqrt{E_t(\hat{\lambda}_t)} = \frac{P_{t,+}}{\sqrt{1+\hat{\lambda}_t}}$. The complementary element, $I - E_t(\hat{\lambda}_t) = P_{t,-} + \frac{\hat{\lambda}_t}{1+\hat{\lambda}_t} P_{t,+}$, can be associated with two measurement operators: $M_2 = P_{t,-}$ and $M_3 = \sqrt{\frac{\hat{\lambda}_t}{1+\hat{\lambda}_t}} P_{t,+}$. In total, we have three measurement operators, $\{M_1, M_2, M_3\}$. We associate them with two different operations on an input sample qubit: the identity $I$ (associated with the outcome of $E_t(\hat{\lambda}_t)$) and the bit-flip $X$ (associated with the outcome of $I - E_t(\hat{\lambda}_t)$). This allows us to construct the three Kraus operators for the channel. According to the Stinespring dilation theorem, this can be realized by introducing two ancilla qubits, which provide a 4-dimensional Hilbert space.

After passing through this channel, the reduced density matrix of the sample qubit is prepared into our desired input state. A direct calculation verifies that the probability of subsequently measuring the sample qubit in the $\ket{0}$ state is:
\begin{equation}
\label{eq:recording_qubit_prob}
p_0(\lambda_t, \hat{\lambda}_t) = \frac{1}{2} \left( 1 + \frac{\lambda_t - \hat{\lambda}_t}{1 + |\hat{\lambda}_t|} \right).
\end{equation}
By comparing this probability with the form of our standard input state, $\bra{0}\rho_{\text{in}}\ket{0} = 1/2+\varepsilon^*$, we can establish the precise relationship between the encoded signal strength $\varepsilon^*$ and the physical deviation:
\begin{equation}
    \varepsilon^* = \frac{\lambda_t - \hat{\lambda}_t}{2(1 + |\hat{\lambda}_t|)}.
\end{equation}

\subsection{The Double-Stage Algorithm}
\label{subapp:double-stage}

Implementing this algorithm requires the introduction of the reverse controlled-overwrite channel, $\mathcal{C}_{\text{r-write}}$, which is described by a set of three Kraus operators:
\begin{align*}
    K_1 &= \ket{1}\bra{1}_{M_1} \otimes I_{M_2}, \\
    K_{0,0} &= \ket{0}\bra{0}_{M_1} \otimes \ket{1}\bra{0}_{M_2}, \\
    K_{0,1} &= \ket{0}\bra{0}_{M_1} \otimes \ket{1}\bra{1}_{M_2}.
\end{align*}
As can be seen, its logic is the inverse of the controlled-overwrite channel $\mathcal{C}_{\text{write}}$: if $M_1$ is in the $\ket{0}$ state, $M_2$ is overwritten to $\ket{1}$; otherwise, its state remains unchanged. Based on this, we present the double-stage algorithm based on the weakly-driven quantum walk, with its detailed procedure shown in Algorithm~\ref{alg:double_stage}.

\begin{algorithm}[H]
\caption{The Double-Stage Algorithm for Learning Pauli Channels}
\label{alg:double_stage}
\begin{algorithmic}[1]
\Resources \parbox[t]{0.85\linewidth}{
    Access to the channel that prepares $\rho_{\text{in}}(\varepsilon^*)$ \\
    A sample qubit $Q_S$ \\
    A pointer qubit $P$ \\
    A first recorder qubit $M_1$ \\
    A second recorder qubit $M_2$
}
\Statex
\Require Parameters $n$ and $\varepsilon$.
\Ensure The final state of the second recorder qubit $M_2$.
\Statex
\State \textbf{Parameter setting:} $m = O(\log n / \varepsilon^2)$, $\theta = O(\varepsilon^2 / \log n)$.
\State \textbf{Initialization:} $M_2 \to \ket{0}\bra{0}$.
\Statex
\For{iteration $k = 1, \dots, 3n$}
    \State $M_1 \to \ket{0}\bra{0}$.
    \For{round $i = 1, \dots, m$}
        \State $P \to \ket{0}\bra{0}$.
        \For{step $j = 1, \dots, i$}
            \State Prepare $\rho_{\text{in}}(\varepsilon^*)$ on $Q_S$.
            \State Apply $\mathcal{C}_{\text{walk}}^{\text{w}}$ on $Q_S$ and $P$.
        \EndFor
        \State Apply $\mathcal{C}_{\text{write}}$ on $P$ and $M_1$.
    \EndFor
    \State Apply $\mathcal{C}_{\text{r-write}}$ on $M_1$ and $M_2$.
\EndFor
\end{algorithmic}
\end{algorithm}

The operational procedure can be described in two stages:
\begin{enumerate}
    \item \textbf{First stage:} We first run the $m$-round weakly-driven quantum walk subroutine, with its result recorded on the first recorder qubit, $M_1$. By selecting the parameters $m=O(\log n / \varepsilon^2)$ and $\theta=O(\varepsilon^2 / \log n)$, we can ensure that the survival probability of $M_1$, denoted $S^{(1)}$, satisfies: $S_0^{(1)} \ge 1/2$ under $H_0$, and $S_1^{(1)} \le 1/n$ under $H_1$.
    \item \textbf{Second stage:} Next, through the $\mathcal{C}_{\text{r-write}}$ operation acting on $M_1$ and $M_2$, we logically invert the probability information from $M_1$ and transfer it to $M_2$. The survival probability of $M_2$ in a single iteration becomes: $1 - S_0^{(1)} \le 1/2$ under $H_0$, and $1 - S_1^{(1)} \ge 1 - 1/n$ under $H_1$. We treat this entire process as a whole and repeat it $3n$ times. In each iteration, $M_1$ is reset, but the results are accumulated in the same qubit $M_2$. The final overall survival probability of $M_2$ is the product of the single-iteration survival probabilities:
    \begin{itemize}
        \item Under $H_0$: The overall survival probability is $S_0^{(2)} \le (1/2)^{3n} = 1/8^n$.
        \item Under $H_1$: The overall survival probability is $S_1^{(2)} \ge (1-1/n)^{3n} \approx e^{-3} = O(1)$.
    \end{itemize}
\end{enumerate}
The total number of channel queries for this scheme is $3n \times (\sum_{i=1}^{m} i) = O(n \cdot m^2) = O(n (\log n / \varepsilon^2)^2) = O(n \log^2 n / \varepsilon^4)$.

\subsection{Performance Analysis in Pauli Channel Estimation}
\label{subapp:pauli_performance}

In the protocol of Ref.~\cite{chenEfficient2025}, the output of the double-stage algorithm, the second recorder qubit $M_2$, is used as a control qubit during the tests of the $4^n-1$ eigenvalues. It acts via another reverse controlled-overwrite channel, $\mathcal{C}_{\text{r-write}}$, on a global recorder qubit, $M_3$, which is not reset throughout the $4^n-1$ tests. Suppose that in the $t$-th test, the survival probability of $M_2$ is $S_t^{(2)}$. Then, the probability $f_t$ that $M_3$ is overwritten to $\ket{1}$ in this test is equal to $S_t^{(2)}$. After all $4^n-1$ eigenvalues have been tested, the final survival probability of $M_3$ is:
\begin{equation}
    S^{(3)} = \prod_{t=1}^{4^n-1} (1 - f_t) = \prod_{t=1}^{4^n-1} (1 - S_t^{(2)}).
\end{equation}
Finally, by measuring $M_3$ and repeating the entire process $O(1)$ times to obtain an estimate of its overwrite probability, we can make a judgment between the following two hypotheses regarding the unknown channel:
\begin{itemize}
    \item \textbf{Hypothesis (1):} The difference $|\lambda_t - \hat{\lambda}_t|$ between all true eigenvalues $\lambda_t$ and their hypothesized values $\hat{\lambda}_t$ is less than an outer threshold $\varepsilon_p$.
    \item \textbf{Hypothesis (2):} There is at least one true eigenvalue $\lambda_{t^*}$ for which the difference $|\lambda_{t^*} - \hat{\lambda}_{t^*}|$ is greater than an inner threshold $\varepsilon'_p = \Theta(\varepsilon_p/\sqrt{\log n})$.
\end{itemize}

First, we show that our algorithm can detect significant signals. Assume that only one eigenvalue, $t^*$, has a significant deviation, i.e., $|\lambda_{t^*} - \hat{\lambda}_{t^*}| \ge \varepsilon_p$, while the differences for all other $4^n-2$ eigenvalues are strictly zero. In this case, the final survival probability of $M_3$ is:
\begin{equation*}
\begin{aligned}
    S^{(3)} &= (1 - S_{t^*}^{(2)}) \cdot \prod_{t \ne t^*} (1 - S_t^{(2)}) \\
    &\le (1 - S_1^{(2)}) \cdot \prod_{t \ne t^*} (1 - S_0^{(2)}).
\end{aligned}
\end{equation*}
Substituting the performance of our double-stage algorithm, $S_1^{(2)} \ge e^{-3}$, we get:
\begin{equation}
\begin{aligned}
    S^{(3)} &\le (1 - e^{-3}) \cdot (1 - S_0^{(2)})^{4^n-2} \\
    &\le (1 - e^{-3}).
\end{aligned}
\end{equation}
This means that as long as there is one significant signal, the probability of $M_3$ being overwritten is at least the constant $e^{-3}$. Therefore, if the measured overwrite probability of $M_3$ is less than $e^{-3}$, we can conclude Hypothesis (1).

Second, we show that our algorithm does not falsely report weak signals. Consider the case where all $4^n-1$ eigenvalues have a non-zero deviation equal to the inner threshold, i.e., $|\lambda_t - \hat{\lambda}_t| = \varepsilon'_p = \Theta(\varepsilon_p/\sqrt{\log n})$. This represents a scenario with interference from a large number of weak signals. As discussed in Sec.~\ref{sec:apply_pauli}, the threshold of our subroutine is set to $\varepsilon = \varepsilon_p / 4 = \Theta(\varepsilon_p)$, so $\varepsilon' = \Theta(\varepsilon/\sqrt{\log n})$.

We first calculate the survival probability of the first recorder qubit, $S^{(1)}(m, \varepsilon')$. Its logarithmic decay is:
\begin{equation*}
    -\ln S^{(1)}(m, \varepsilon') = \frac{m^2\theta^2}{8} + \frac{m^3\theta^2\varepsilon'^2}{3}.
\end{equation*}
Substituting the parameters $m$, $\theta$, and $\varepsilon'=\Theta(\varepsilon/\sqrt{\log n})$:
\begin{equation*}
\begin{aligned}
    -\ln S^{(1)}(m, \varepsilon') &= O(1) + O\left(\frac{\log^3 n}{\varepsilon^6} \cdot \frac{\varepsilon^4}{\log^2 n} \cdot \frac{\varepsilon^2}{\log n}\right) \\
    &= O(1) + O(1) = O(1).
\end{aligned}
\end{equation*}
This implies that even with interference from weak signals, the survival probability of $M_1$ remains a constant greater than zero. After the logical inversion, the single-iteration survival probability of $M_2$ is $1 - S^{(1)}(m, \varepsilon')$, which is a constant less than one. After $3n$ iterations, the overall survival probability of $M_2$ for any test $t$, $S_t^{(2)}$, becomes $(1 - S^{(1)}(m, \varepsilon'))^{3n} = e^{-\Omega(n)}$.

Therefore, the final survival probability of $M_3$ is:
\begin{equation}
\begin{aligned}
    S^{(3)} &= \prod_{t=1}^{4^n-1} (1 - S_t^{(2)}) = \left(1 - e^{-\Omega(n)}\right)^{4^n-1} \\
    &\approx 1 - (4^n-1)e^{-\Omega(n)}.
\end{aligned}
\end{equation}
As long as we choose the constant factors such that the decay rate of $e^{-\Omega(n)}$ is faster than the growth rate of $4^n$, the final overwrite probability of $M_3$ will be exponentially small in $n$. Thus, if the measured overwrite probability of $M_3$ is greater than or equal to $e^{-3}$, we can conclude Hypothesis (2).

\bibliography{main}

\end{document}